\documentclass[twocolumn,aps,prl,superscriptaddress,nofootinbib]{revtex4-1}
\usepackage{graphicx}
\usepackage{subcaption}
\usepackage{color}
\graphicspath{{figures/}{fig/}}
\usepackage{amsmath}
\usepackage{amssymb}
\usepackage{bm}
\usepackage{slashed}
\usepackage{epsfig}
\usepackage{amsfonts}
\usepackage{epstopdf}
\usepackage{hyperref}
\usepackage{bbm}
\usepackage{textcomp}
\usepackage{color}
\newcommand{\sect}[1]{{\it \textbf{#1.} --- }}
\newcommand{\NOdisplay}[1]{ }
\def\MSbar{\overline{\mathrm{MS}}}
\def\as{\alpha_s}
\def\Gamt{\mathrm{\Gamma}_t}
\def\HFs{f_{\mathrm{L,R,0}}}
\def\HFp{f_{\mathrm{L}}}
\def\HFm{f_{\mathrm{R}}}
\def\HFl{f_{0}}
\def\HTensor{\mathcal{W}_{tb}^{\,\mu \nu }}
\def\TbWprocess{t \rightarrow b + W^+ + X_{\mathrm{\tiny QCD}}}
\begin{document}
\title{Top-Quark Decay at Next-to-Next-to-Next-to-Leading Order in QCD}

\author{Long Chen}
\email{longchen@sdu.edu.cn}
\affiliation{School of Physics, Shandong University, Jinan 250100, China}

\author{Xiang Chen}
\email{xchenphy@pku.edu.cn}
\affiliation{School of Physics, Peking University, Beijing 100871, China}

\author{Xin Guan}
\email{guanxin0507@pku.edu.cn}
\affiliation{School of Physics, Peking University, Beijing 100871, China}

\author{Yan-Qing Ma}
\email{yqma@pku.edu.cn}
\affiliation{School of Physics, Peking University, Beijing 100871, China}
\affiliation{Center for High Energy Physics, Peking University, Beijing 100871, China}

\date{\today}

\begin{abstract}
We present the first complete high-precision QCD corrections to the inclusive decay width $\Gamt$, the $W$-helicity fractions $\HFs$ and semi-inclusive distributions for the top-quark decay process $\TbWprocess$ at NNNLO in the strong coupling constant $\as$.
In particular, the pure NNNLO QCD correction decreases the $\Gamt$ by about $0.8\%$ of the previous NNLO result at the top-quark pole mass scale, exceeding the error estimated by the usual scale-variation prescription. 
After taking into account all sources of errors, we get $\Gamt = 1.3148^{+0.003}_{-0.005} \times|V_{tb}|^2+ 0.027\,(m_t - 172.69)\,\text{GeV} $, the error of which meets the request by future colliders.
On the other hand, the NNNLO QCD effects on $\HFs$ are found to be much smaller, at the level of one per-mille for the dominating $\HFl$, predestining them to act as precision observables for the top-quark decay process.
\end{abstract}

\maketitle
\allowdisplaybreaks


\sect{Introduction}
The top-quark $t$, to-date heaviest fundamental particle ever discovered in experiments, plays a special role both in the precision test of the Standard Model (SM), especially the electroweak sector, as well as in searching for New Physics.
The properties of its production and decay have been actively studied at the Tevatron and LHC, and will also be among the core physical programs at future high-energy lepton colliders~\cite{ILC:2013jhg,CLICdp:2018esa,FCC:2018evy,CEPCStudyGroup:2018ghi}.

The large $t$-quark mass $m_t$, currently measured to be $172.69 \pm 0.30\,$GeV\cite{ParticleDataGroup:2022pth} usually interpreted as the pole mass, not only sets a sufficiently-large scale to apply perturbative Quantum Chromodynamics (QCD), but also makes for a unique feature relevant for QCD phenomenology: the $t$-quark decays almost exclusively to $W$ and bottom-quark $b$ before it hadronizes. 
The current most precise measurement for the $t$-quark decay width $\Gamt$ comes from the CMS \cite{CMS:2014mxl} which gives $1.36\pm0.02\mathrm{(stat.)}^{+0.14}_{-0.11}\mathrm{(syst.)}\,$GeV.
The anticipated experimental uncertainties in the measurement of $\Gamt$ at the future hadron and lepton colliders can be reduced to about $20 \sim 26\,$MeV\cite{Horiguchi:2013wra,CLICdp:2018esa,Baskakov:2018huw,Li:2022iav}. To fully take advantage of the data, the theoretical error should be at least smaller than one-third of the experimental one, say less than  $7\,$MeV.
Although QCD corrections to $t$-quark decay have been calculated perturbatively up to next-to-next-to leading order (NNLO) in the strong coupling $\as$, e.g.~in refs.\cite{Jezabek:1988iv,Czarnecki:1990kv,Li:1990qf,Czarnecki:1998qc,Chetyrkin:1999ju,Fischer:2001gp,Blokland:2004ye,Blokland:2005vq,Czarnecki:2010gb,Gao:2012ja,Brucherseifer:2013iv,Campbell:2020fhf,Meng:2022htg,Chen:2022wit}, we will see later that the theoretical error at NNLO cannot meet this request.
Besides, there are concerns over the convergence rate of the perturbative series related to the fact that both the pole mass $m_t$ and the decay width $\Gamt$ are sensitive to the infrared-renormalon issue (See~e.g.~refs.\cite{Beneke:1994qe,Smith:1996xz,Beneke:1998ui,Beneke:2016cbu,Hoang:2017btd,FerrarioRavasio:2018ubr}).
It is thus desirable to explicitly determine the complete QCD corrections at next-to-next-to-next-to leading order (NNNLO) in $\as$, which we accomplish for the first time in this Letter.

One particularly interesting phenomenon in $t$-quark decay is that the produced $W$ is polarized even if the $t$-quark is unpolarized, due to the chiral structure of the weak interaction. 
The current measurements~\cite{ATLAS:2022rms} of fractions of longitudinal ($f_{0}$), left-handed ($f_{\mathrm{L}}$) and right-handed ($f_{\mathrm{R}}$) polarization states of $W$ at the ATLAS are found to be $f_{0} = 0.684 \pm 0.005\,\mathrm{(stat.)}  \pm 0.014\,\mathrm{(syst.)}$, $f_{\mathrm{L}} = 0.318 \pm 0.003\,\mathrm{(stat.)} \pm 0.008\,\mathrm{(syst.)}$ and $f_{\mathrm{R}} = -0.002 \pm 0.002\,\mathrm{(stat.)} \pm 0.014\,\mathrm{(syst.)}$.
The current uncertainties in $\HFl$ and $\HFp$ are dominated by  
the $t\bar{t}$-production modeling, jet energy scale and also 
the experimental uncertainties in $m_t$, while for $\HFm$ the uncertainties in $\as$ and $b$-quark mass $m_b$ matter too. 
All these experimental uncertainties are expected to be significantly improved at future lepton colliders~\cite{ILC:2013jhg,CLICdp:2018esa,FCC:2018evy,CEPCStudyGroup:2018ghi,Vos:2016til,Schwienhorst:2022yqu}, and the experimental uncertainties in $f_{\mathrm{L,0}}$ could therefore be comparable or smaller than the theoretical errors based on NNLO QCD calculations~\cite{Czarnecki:2010gb,Gao:2012ja,Brucherseifer:2013iv}.
It is thus also desirable to improve the accuracy of the theoretical predictions for these precision observables at higher orders in QCD.

In this Letter we formulate an efficient method to tackle this problem, which not only allows us to obtain high-precision results at one order higher in QCD but also provides us with the ingredients that would be needed to arrive at a fully differential calculation of any infrared-safe observables for $t$-quark decay process at NNNLO later on.

\sect{The method}
Aiming for obtaining NNNLO QCD corrections to the semi-inclusive distributions related to the $W$ produced in $\TbWprocess$, from which inclusive observables such as $\Gamt$ and $\HFs$ can be composed as well, we write $\Gamt$ in terms of the semi-inclusive hadronic tensor $\mathcal{W}^{\,\mu\nu}_{tb}$ integrated over the $W$ momentum $k$ as follows,%
{\small
\begin{equation}
\label{eq:SemiDiffDW}    
\Gamt =\frac{1}{2\, m_t}\, \int \frac{\mathrm{d}^{d-1} k}{(2 \pi )^{d-1} 2 E}\, \HTensor 
\sum_{\lambda}^{L,R,0}\, \varepsilon^{*}_{\mu} (k, \lambda)\, \varepsilon_{\nu} (k, \lambda)\,,
\end{equation}
}%
with fixed $t$-quark momentum $p$.
For an on-shell $W$ with mass $m_W$, the polarization-sum $\sum_{\lambda}\, \varepsilon^{*}_{\mu} (k, \lambda)\, \varepsilon_{\nu} (k, \lambda)$ can be reduced to $\big(g^{\mu\nu} - k^{\mu} k^{\nu}/m_W^2\big)$. For $W$ with a definite helicity $\lambda$, the corresponding projectors can be found in ref.~\cite{Fischer:2000kx} and see also ref.~\cite{Chen:2019wyb}.
Anticipating potential infrared-soft and/or collinear (IR) divergence in the phase-space integration over $k$ in the region where $E$ reaches its maximum, we have introduced the dimensional regularization (DR) with spacetime dimension denoted by $d \equiv 4 - 2\epsilon$.

$\HTensor$ can be decomposed to 5 linearly-independent Lorentz-tensor structures:%
{\small
\begin{eqnarray}
\label{eq:HTffs}    
\mathcal{W}_{tb}^{\,\mu \nu }(p, k) &=& 
W_1 \, g^{\mu\nu} \,+ \,
W_2 \, p^{\mu} p^{\nu} + \,
W_3 \, k^{\mu} k^{\nu}  \nonumber\\
&+& W_4 \,  \big(p^{\mu} k^{\nu} \, +\, k^{\mu} p^{\nu} \big) \,+\,
W_5 \, i\epsilon^{\mu \nu \rho \sigma}\, p_\rho \, k_\sigma \, ,
\end{eqnarray}
}%
where the Levi-Civita tensor $\epsilon^{\mu \nu \rho \sigma}$ appears due to the chiral structure of the weak $tbW$-vertex.
Since the $\gamma_5$ from the $tbW$-vertex always appears on an open fermion chain of the contributing QCD amplitudes, $\gamma_5$ can be treated fully  anticommutively~\cite{Bardeen:1972vi,Chanowitz:1979zu,Gottlieb:1979ix,Korner:1991sx} in a straightforward manner.

Each form factor $W_i$ is a function of $m_t,\, m_W$ and the $W$-energy $E$, as well as $m_b$, and receives both virtual and real-radiation type QCD corrections.
Loop integrals are reduced to relatively simpler master integrals using integration-by-parts (IBP) identities~\cite{Chetyrkin:1981qh}, employing the recently released package {\tt Blade}~\cite{Guan:2024byi}, which is armed with the strategy of block-triangular form\cite{Liu:2018dmc,Guan:2019bcx} and utilizes the {\tt FiniteFlow}~\cite{Peraro:2019cjj} package
(For other reduction packages on the market, see refs.~\cite{Anastasiou:2004vj, Smirnov:2008iw,Smirnov:2013dia,Smirnov:2014hma,Smirnov:2019qkx, Maierhofer:2017gsa, Maierhofer:2018gpa,Klappert:2020nbg, Lee:2012cn, Lee:2013mka, Studerus:2009ye,vonManteuffel:2012np, Wu:2023upw}).
The resulting master integrals are computed using the differential equation method~\cite{Kotikov:1990kg,Remiddi:1997ny} based on series expansion (See~e.g.~refs.\cite{Caffo:2008aw,Czakon:2008zk,Lee:2014ioa, Moriello:2019yhu, Hidding:2020ytt, Armadillo:2022ugh}). 
To fix the boundary conditions for these differential equations, we utilize the auxiliary mass flow method~\cite{Liu:2017jxz,Liu:2020kpc,Liu:2021wks,Liu:2022mfb} implemented in {\tt AMFlow}~\cite{Liu:2022chg}.
The phase-space integrals over the momenta of final-state particles, except for $k$ being fixed, are treated in the same manner as loop integrals by means of the reverse unitarity~\cite{Anastasiou:2002yz,Anastasiou:2002qz,Anastasiou:2003yy}. 
To give an idea of the complexity encountered in our calculation, the number of integrals in the $\mathcal{O}(\as^3)$ corrections to $\HTensor$ is about $7\times 10^4$, which are reduced to 2988 master integrals to be solved using the aforementioned method. 
Armed with all these highly efficient techniques, we are able to construct a piecewise series expansion representation (PSE) for each master integral, deeply expanded up to about 200 orders in $E$, using the differential equation solver in {\tt AMFlow}.
Consequently, a high-precision result for $\HTensor$ is obtained in the form of deeply-expanded PSE.

Here comes a tricky point when completing the phase-space integration of $\HTensor$ over $k$ for calculating physical observables: 
$W_i$ in \eqref{eq:HTffs} contain IR divergence in the region with $E$ reaching its maximum, and the integration over $k$ shall be done with proper regularization which we take to be DR. 
The phase-space integration of $\HTensor$ over the IR-dangerous regions is done with $\epsilon$ assigned with the values, $10^{-3} + n \times 10^{-4}$ for $n=0,~1,~\cdots, 15$, to regularize the potential IR divergences~\cite{Liu:2022mfb,Liu:2022chg}.
The fit regarding the $\epsilon$-dependence is done only at the very end for the final finite physical objects of interest, which can be the inclusive $\Gamt$ and $\HFs$ as well as IR-safe distributions.
This treatment saves us from the need to explicitly perform Laurent expansions in $\epsilon$ during the intermediate stages, which significantly reduces the computational time for our problem.
This method is anticipated to work also with physical jet-based observables where the dimensionally-regularized phase-space can be conducted using numerical integration methods such as the Monte Carlo technique, with certain phase-space cuts imposed. (The numerical values for $\epsilon$ shall be adapted.)
We validated the above method using the simplest inclusive observable of the process, the $\Gamt$, by comparing the numerical results composed from $\HTensor$, to be presented below, against those calculated by applying the optical theorem (where the integration over $k$ can be done using the reversed unitarity~\cite{Anastasiou:2002yz,Anastasiou:2002qz,Anastasiou:2003yy}). We find a perfect agreement between the two methods within our high numerical precision.

\sect{The inclusive decay width}
We are now ready to present our numerical results and begin with $\Gamt$.
The QCD effects on $\Gamt$ in SM can be parameterized as%
\begin{equation}
\label{eq:parametricDW}
\Gamt = \Gamma_0\, \Big[ \mathbf{c}_0 + \frac{\as}{\pi}  \mathbf{c}_1 + \Big(\frac{\as}{\pi}\Big)^2 \mathbf{c}_2 + \Big(\frac{\as}{\pi}\Big)^3  \mathbf{c}_3 
+ \mathcal{O}(\as^4) \Big] \, ,
\end{equation}%
where we introduced a prefactor 
$\Gamma_0\equiv\frac{G_F\,\, m_W^2\, m_t \, |V_{tb}|^2}{12 \sqrt{2} }$ 
with $G_F$ the Fermi constant, the CKM matrix element $V_{tb}$ taken to be 1 in the following numerical results.
The perturbative coefficients $\mathbf{c}_i$ are functions of the kinematic ratio $m_W^2/m_t^2$ in the limit $m_b = 0$, and also the renormalization scale $\mu$ in general.
Below we present the numerical results explicitly in a scheme where $m_t$ is renormalized in the on-shell scheme and $\as$ is $\MSbar$-renormalized with 5 massless quark flavors.

At the scale $\mu = m_t/2$, close to the kinetic energy $m_t - m_W - m_b$ of the final state,  we obtain %
\begin{eqnarray}
\label{eq:ciV}
&& \mathbf{c}_0 = 1.93851\,,  
\quad \mathbf{c}_1 = -4.85519\,, \nonumber\\
&& \mathbf{c}_2 = -21.2260 \,, ~
\mathbf{c}_3 = -174.265\,, 
\end{eqnarray}%
using the SM input parameters $m_t = 172.69$ GeV and $m_W = 80.377$ GeV.  
In \eqref{eq:ciV} we have truncated our internal high-precision results for the perturbative coefficients $\mathbf{c}_i$ to the first 6 digits for the sake of display (Results with higher precision can be obtained from the supplementary file associated with this Letter.)
Up to NNLO our result~\eqref{eq:ciV} agrees with those in refs.~\cite{Blokland:2004ye,Blokland:2005vq,Brucherseifer:2013iv,Chen:2022wit}.
With $G_F = 1.166379\times 10^{-5} $ GeV$^{-2}$ and $\alpha_s(m_t/2) \approx 0.1189$ at scale $\mu=m_t/2$, obtained by solving the renormalization-group equation for the running $\as$ at four-loop order~\cite{vanRitbergen:1997va,Chetyrkin:1999pq,Czakon:2004bu,Chetyrkin:2004mf} with input $\alpha_s(m_Z) = 0.1179$ at the Z-pole mass $m_Z = 91.1876$ GeV, we further obtain%
{\small 
\begin{eqnarray}
\label{eq:explicitDW}
\Gamt &=& 1.48642 -0.140877 -0.023306 -0.007240 \,\,\text{GeV}  \nonumber\\
  &=& 1.31500\,\,\text{GeV} \, ,   
\end{eqnarray}
}%
where the first line provides the decomposition of the total NNNLO result according to the $\as$ order. 
Therefore the QCD corrections continue to decrease the Born-level result for $\Gamt$ up to $\mathcal{O}(\as^3)$. 

\begin{figure}[htbp]
\includegraphics[scale=0.8]{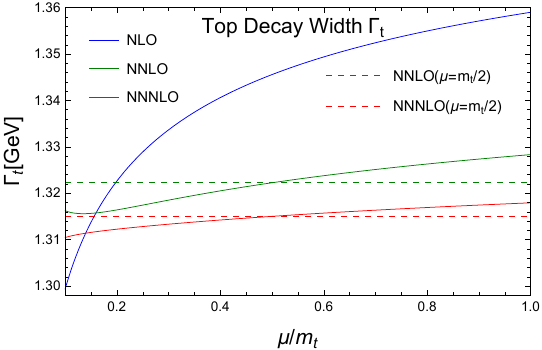}
\caption{The scale dependence of the fixed-order results for $\Gamt$ in $\mu/m_t \in [0.1, 1]$}
\label{fig:TotWidthMuD}
\end{figure}

Results for $\Gamt$ at other scales can be readily derived from \eqref{eq:explicitDW} using the renormalization-group equation method, which are provided as a function in the supplementary file where the $m_t$ value can be changed as well. 
In FIG.~\ref{fig:TotWidthMuD} we plot the scale dependence of our results for $\Gamt$ for $\mu/m_t \in [0.1, 1]$ up to NNNLO.%
It can be found that the scale dependence has been further improved at NNNLO, as expected. 
However, due to the turning point of the NNLO green curve (at about $\mu/m_t=0.14$) in FIG.~\ref{fig:TotWidthMuD}, its scale variation can never cover the NNNLO result at any scales less than $\mu/m_t=0.6$, including the central value $\mu/m_t=0.5$ chosen in this Letter.
This clearly demonstrates that the NNLO results can underestimate the theoretical error by simply studying the $\mu$ dependence, and thus determining the $\mathcal{O}(\as^3)$ corrections explicitly is very important. We note that at the benchmark scale $\mu = m_t$ typically chosen in literature, the pure $\mathcal{O}(\as^3)$ correction decreases $\Gamt$ by about $0.8\%$ of the previous NNLO result, roughly $10\,$MeV, significantly exceeding the usual NNLO scale uncertainty.

Despite an overall good convergence in \eqref{eq:explicitDW}, by examining the successive ratios $\mathbf{c}_{i+1}/\mathbf{c}_i$ starting from $i=1$, we begin to see the hint that the convergence of the perturbative series seems to deteriorate as the perturbative order goes higher.
This is expected due to the well-known infrared-renormalon sensitivity of the on-shell (pole) mass definition for heavy quarks (see, e.g.~refs.\cite{Beneke:1994qe,Smith:1996xz,Beneke:1998ui,Beneke:2016cbu,Hoang:2017btd,FerrarioRavasio:2018ubr}), which can be avoided when rewritten using the renormalon-free $\MSbar$-mass $\overline{m}_t$. 
Using the three-loop conversion formula from the pole mass $m_t$ to $\overline{m}_t$\cite{Melnikov:2000qh}, we get $\overline{m}_t(\overline{m}_t) = 163.094$ GeV. Then we obtain the corresponding decay rate $\bar{\Gamma}_t(\mu=\overline{m}_t) = 1.2125 \,+\, 0.09830 \,+\, 0.00845 \,+\, 0.00034 = 1.31959$ GeV, the convergence rate of which is clearly improved, as compared to \eqref{eq:explicitDW}.

\sect{The $W$-helicity fractions and angular asymmetries} 
As the first example beyond the inclusive $\Gamt$ at NNNLO, we consider the $W$-helicity fractions $\HFs$ resulting from a $t$-quark decay, unequal between each other due to the chiral structure of the $tbW$-vertex in SM.

To present our results for QCD corrections to $\HFs$, following from their very definitions, we introduce further $f^{[n]}_{\lambda}$ accurate to the $n$-th order in $\as$ as in ref.~\cite{Czarnecki:2010gb},
$f^{[n]}_{\lambda} = \frac{\sum_{i=0}^{n} \Gamma^{[n]}_{\lambda}}{\sum_{i=0}^{n} \Gamma^{[n]}_t}$
where $\Gamma^{[n]}_t = \sum_{\lambda = 0,L,R} \Gamma^{[n]}_{\lambda}$ is the total top-decay width at $\mathcal{O}(\as^n)$, expressed as a sum over partial width $\Gamma^{[n]}_{\lambda}$ with a polarized $W$ defined by 
$\Gamma^{[n]}_{\lambda} = \frac{1}{2\, m_t}\int \frac{\mathrm{d}^{d-1} k}{(2 \pi )^{d-1} 2 E}\, \HTensor \varepsilon^{*}_{\mu} (k, \lambda)\, \varepsilon_{\nu} (k, \lambda)$ at $\mathcal{O}(\as^n)$ with the projector for $\varepsilon^{*}_{\mu} (k, \lambda)\, \varepsilon_{\nu} (k, \lambda)$ taken from  ref.~\cite{Fischer:2000kx}.

Choosing our default parameters as used above, we obtain the following numerical results for the NNNLO QCD corrections to the unexpanded ratios:%
{\small
\begin{eqnarray}
\label{eq:HFres_0m}
\HFl^{[3]} &=& 0.697706 -0.008401 -0.001954 -0.000613 \,,\nonumber\\
&=& 0.686737\,,\nonumber\\
\HFp^{[3]} &=& 0.302294 + 0.007254+ 0.001799 + 0.000586 \,,\nonumber\\
&=& 0.311933 \,,\nonumber\\
\HFm^{[3]} &=& 0. + 0.001147 + 0.000155 + 0.000027 \,,\nonumber\\
&=& 0.001330 \,,
\end{eqnarray}
}%
at $\mu=m_t/2$, where the numbers in the first equality for each $f^{[3]}_{\lambda}$ give the difference compared to the proceeding perturbative order.
\footnote{Alternatively, one may expand $f^{[3]}_{\lambda}$ in $\alpha_s$ and truncate to the third order, resulting $\HFl^{[3]} = 0.686941\,,\, \HFp^{[3]} = 0.311747\,,\, \HFm^{[3]} = 0.001312 \,$. The differences compared to the unexpanded results \eqref{eq:HFres_0m} are much smaller than the QCD scale uncertainty of $\Gamma^{[3]}_t$ (except for the tiny $\HFm^{[3]}$).}
When evaluated at $\mu=m_t$, our results up to NNLO are in good agreement with the numbers given in ref.~\cite{Czarnecki:2010gb}. 
Defined as ratios, the scale uncertainties of $\HFs$ are naturally quite small.

The three $\HFs$ with numerical results in \eqref{eq:HFres_0m} govern the  $\cos{\theta^{*}}$ angular distribution through~\cite{Aguilar-Saavedra:2015yza}%
{\small 
\begin{equation*}
\label{eq:angDistInHFs}
\frac{1}{\Gamt}\frac{\mathrm{d} \Gamt}{\mathrm{d} \cos\theta^*} 
= \frac{3}{4}(\sin^2\theta^*) \,f_0 +  
\frac{3}{8}(1-\cos\theta^*)^2\, f_\mathrm{L} + 
\frac{3}{8}(1+\cos \theta^*)^2\, f_\mathrm{R}\,,
\end{equation*}
}%
where $\theta^{*}$ is the angle between the momentum of the charged lepton from the $W$-decay in $W$-rest frame and the $W$-momentum in $t$-rest frame.%
\NOdisplay{
In FIG.~\ref{fig:angDistribution} we plot our results for the $\cos{\theta^{*}}$ angular distribution up to NNNLO in QCD, confined to the window $\pi/4 < \theta^{*} < 3\pi/4$ for a better display.%
\begin{figure}[htbp]
\includegraphics[scale=0.6]{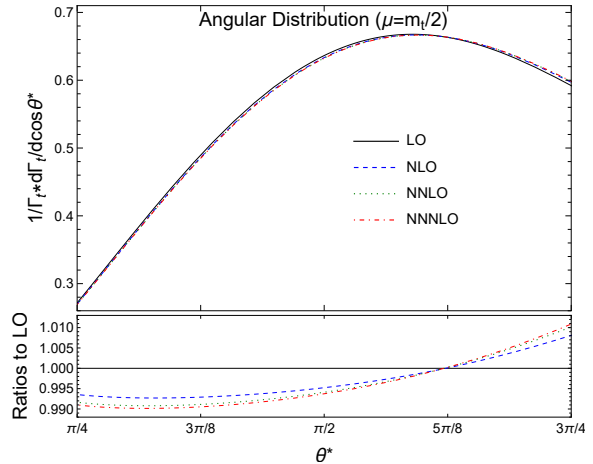}
\caption{Angular distribution of the charged lepton from $t$-quark decay in the $W$ rest frame.}
\label{fig:angDistribution}
\end{figure}
For reducing measurement uncertainties when extracting results on $\HFs$, ref.~\cite{Aguilar-Saavedra:2006qvv} suggested a generalized angular asymmetry based on this $\cos{\theta^{*}}$ distribution, 
{\small
\begin{equation*}
\mathcal{A}_z = \frac{\mathrm{N}(\cos{\theta^{*}} > z) - \mathrm{N}(\cos{\theta^{*}} < z)}{\mathrm{N}(\cos{\theta^{*}} > z) + \mathrm{N}(\cos{\theta^{*}} < z)}     ,
\end{equation*}
}%
where $\mathrm{N}(\cos{\theta^{*}} > z) = 
\int_{z}^{1} \frac{\mathrm{d} \Gamt}{\mathrm{d} \cos\theta^*}\, \mathrm{d} \cos\theta^*$, proportional to the number of charged leptons in $t$-quark decay observed with $\cos{\theta^{*}}$ larger than a given value $z$ (and similarly for $\mathrm{N}(\cos{\theta^{*}} < z)$).}%
The forward-backward asymmetry of the charged lepton in $\cos{\theta^{*}}$ distribution corresponds to the special case of $\mathcal{A}_z$ at $z=0$, which equals to $\frac{3}{4}\big(\HFp - \HFm\big)$.
Our results \eqref{eq:HFres_0m} can be readily used to determine the QCD corrections to these quantities up to NNNLO.

\begin{figure}[htbp]
\includegraphics[scale=0.6]{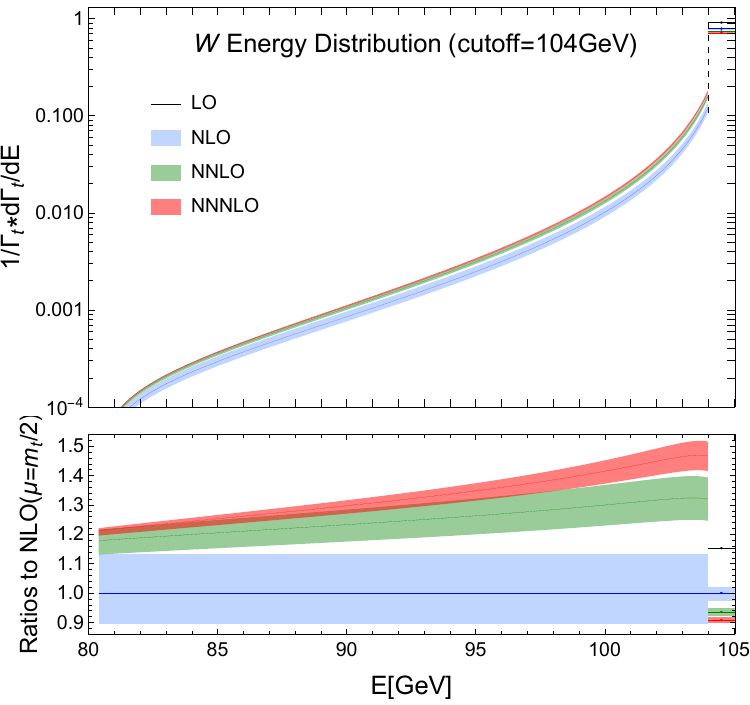}
\caption{The $W$-energy distribution in $t$-quark decay observed in the $t$-quark rest frame.}
\label{fig:WenDistribution}
\end{figure}

\sect{The $W$-energy distribution}
Having the results for $\HTensor$, we are able to calculate the NNNLO QCD corrections to the $W$-energy distribution in $t$-quark decay observed in the $t$-quark rest frame, shown in FIG.~\ref{fig:WenDistribution}.
The distribution of $W$-energy $E$ increases as $E$ becomes larger, and becomes singular in the limit of $E$ reaching its maximum $E_{\mathrm{max}}$ where the QCD radiations have to be either soft and/or collinear to the $b$-quark.
The fixed-order prediction breaks down in this region, and we take an average over a $1\,$GeV-bin in the rightmost end of the distribution, where the QCD corrections up to $\mathcal{O}(\as^3)$ decrease the Born-level result in a way similar to the $\Gamt$ case. 

From the lower panel of FIG.~\ref{fig:WenDistribution}, one sees that the QCD corrections are positive and quite sizable, in particular, the pure $\mathcal{O}(\as^3)$ correction modifies the lowest order result by about $7\sim 14\%$ for $E \in [94, 104]\,$GeV.
The scale variation of the next-to-leading (NLO) curve is solely determined by the change of $\as$, and thus the blue band is independent of $E$. Starting from NNLO, the appearance of large logarithm $\ln\big((E_{\mathrm{max}} - E) /E\big)$ leads to the increase of corrections as $E$ approaches $E_{\mathrm{max}}$.
At NNNLO, the change in the scale-variation band in this region becomes more visible due to poor convergence of the perturbative series contaminated by the aforementioned large logarithmic structure.

\sect{Miscellaneous effects}
With the expression for $\Gamt$ as a deeply-expanded PSE  w.r.t. $m_W^2/m_t^2$ at our hand, valid in the whole physical region, we can investigate the so-called off-shell-$W$ effect up to $\mathcal{O}(\as^3)$.
The standard definition of $\tilde{\Gamma}_t$ with off-shell-$W$ effect (See~e.g.~ref.\cite{Jezabek:1988ja}) is obtained by replacing the $W$ propagator by a Breit-Wigner distribution, where the $W$-momentum square is essentially the invariant-mass of the lepton pair produced from $W$-decay. 
$\tilde{\Gamma}_t$ assumes a similar perturbative parametrization as in \eqref{eq:parametricDW} but with the $i$-th order coefficient denoted as $\tilde{\mathbf{c}}_i$ for distinction.  
With $W$ total decay width $\Gamma_W = 2.085$ GeV, $\delta_i \equiv\big(\tilde{\mathbf{c}}_i-\mathbf{c}_i\big)/\mathbf{c}_i$ are found to be quite small and decrease, albeit very slowly, as the $\as$-order increases:
$\delta_i$ takes $-1.54\%, -1.53\%, -1.39\%, -1.23\% $ respectively for $i=0,1,2,3$.

For the finite $b$-quark mass effect, we denote the $i$-th order coefficient for $\Gamt$ with a non-zero $m_b$ as $\mathbf{c}^{m_b}_i$. 
With $m_b = 4.78\,$GeV we find that $\big(\mathbf{c}^{m_b}_1-\mathbf{c}_1\big)/\mathbf{c}_1 \approx \big(\mathbf{c}^{m_b}_2-\mathbf{c}_2\big)/\mathbf{c}_2 \approx -1.47 \%$.
This strongly indicates that the small non-zero $m_b$ effect at $\mathcal{O}(\as^3)$ may observe a similar small ratio, well-below sub-per-mille level for the total $\Gamt$.

Taking into account the aforementioned finite $m_b$ and off-shell $W$ effects, as well as the NLO electroweak corrections~\cite{Denner:1990cpz,Denner:1990ns,Eilam:1991iz,Do:2002ky,Basso:2015gca} which we re-derived and included by a multiplicative $K$-factor\footnote{
To be  more specific, the central value in~\eqref{eq:explicitDWfin} was obtained by multiplying our full NNNLO QCD result for $\Gamt$, with the finite $m_b$ and off-shell W effects, by our re-calculated NLO electroweak $K$-factor $1.0168$ (See also \cite{Denner:1990cpz,Denner:1990ns,Eilam:1991iz,Do:2002ky,Basso:2015gca}).
If these NLO electroweak corrections are included in an additive manner, the  central value then reads $\Gamt = 1.3180\,$GeV, leading to a difference still compatible with the perturbative error given in \eqref{eq:explicitDWfin}.},
our final result for $\Gamt$ reads%
\begin{eqnarray}
\label{eq:explicitDWfin}
\Gamt = 1.3148^{+0.003}_{-0.005} \times|V_{tb}|^2 + 0.027\,(m_t - 172.69)\,\text{GeV},\, 
\end{eqnarray}%
where the second term parameterizes the main source of error on $\Gamt$ originated from the experimental uncertainty of the input $t$-quark mass value, furthermore $V_{tb}$ is restored explicitly for completeness.
The numbers in the super- and sub-script in \eqref{eq:explicitDWfin} correspond to a conservative estimate of the QCD scale uncertainty obtained by varying $\mu/m_t$ in $[0.1, 1]$.

Being ratios, the $m_b$ effects on $\HFs$ are even smaller, below one per-mille, except for the small $\HFm$ whose Born-level expression vanishes at $m_b=0$.
After incorporating the NLO electroweak correction~\cite{Do:2002ky} our final results for $\HFs$ read%
{\small
\begin{align}
\label{eq:HFresfin}
\HFl^{[3]} = 0.686^{+0.002}_{-0.003}\, ,~
\HFp^{[3]} = 0.312^{+0.001}_{-0.002}\, ,~
\HFm^{[3]} = 0.00157^{+0.00002}_{-0.00002}\,.
\end{align}
}%
Since the scale uncertainties of $\HFs$ are naturally quite small due to them being ratios, the errors for the above results, given in numbers in super- and sub-script are obtained by combining the errors of the following main sources: the conservative QCD-scale-uncertainty error for $\Gamt$ determined in~\eqref{eq:explicitDWfin} and the errors induced by the input $t$-quark mass $172.69 \pm 0.30\,$GeV and $b$-quark mass $4.78^{+0.02}_{-0.03}\,$GeV, as well as that of $\alpha_s(m_Z)=0.1179\pm 0.0009\,$\cite{ParticleDataGroup:2022pth} in the case of $\HFm^{[3]}$.

Our results for $\Gamt$ and $\HFs$ in \eqref{eq:explicitDWfin} and \eqref{eq:HFresfin} respectively are thus sufficient to meet the anticipated precision requirement on the theoretical predictions for these observables to be measured at the future hadron and lepton colliders\cite{Horiguchi:2013wra,CLICdp:2018esa,Baskakov:2018huw,Li:2022iav}, as quoted in the Introduction. 

\sect{Conclusion}
We have performed the first complete NNNLO QCD corrections to the $t$-decay width $\Gamt$, $W$-helicity fractions $\HFs$ as well as the related asymmetries and $W$-energy distribution  in this Letter, in which the off-shell-$W$ and finite $m_b$ effects are addressed as well.
At $\mu=m_t/2$, our best prediction for $\Gamt$ reads $\Gamt =1.3148\,$GeV with a conservative QCD scale-uncertainty corresponding to $[+3,-5]\,$MeV.
The QCD effects in $\HFs$ are found to be much smaller and thus the corresponding theoretical errors are quite small. 
For the $W$-energy distribution, we see that it receives quite sizable QCD corrections. 
Moreover, the treatment of the phase-space integration over $W$-momentum in the current computation may be adapted and further optimized to become useful in the fully differential calculation of any infrared-safe observables for $t$-quark decay process at NNNLO in QCD.
These results represent the most precise theoretical predictions to date and will be useful for the purpose of testing the SM and probing New Physics.

The calculations accomplished in this Letter have demonstrated the efficiency of our approach formulated.
It can be readily applied to the decay of polarized $t$-quarks at NNNLO, where phenomenologically more interesting observables can be constructed, as well as to the mixed QCD-electroweak corrections needed to achieve a reliable per-mille level phenomenological analysis for this process.
Last but not least, our approach and results can be equally applied in studies of other heavy-to-light quark decays, in particular the B-meson semi-leptonic decay. 
For example, the lepton-pair invariant-mass spectrum in this process can be readily evaluated using the functions provided in the associated supplementary file.

\begin{acknowledgments}
\sect{Acknowledgments}
The work of L.C. was supported by the Natural Science Foundation of China under contract No.~12205171, No.~12235008 and No.~12321005.
The work of X.C., X.G. and Y.Q.M. were supported in part by the National Natural Science Foundation of China (Grants No.~12325503, No.~11975029, No.~11875071), the National Key Research and Development Program of China under Contracts No.~2020YFA0406400. 
We acknowledge the computational support from the High-performance Computing Platform of Shandong University and Peking University.

\textbf{Note added}: While the work in this Letter was being finalized, a preprint~\cite{Datta:2023otd} appeared in which the three-loop QCD correction to heavy-light form factors was obtained in the color-planar limit. 
Parallel to the calculations done in this Letter, the leading-color part of the $t$-quark inclusive decay width $\Gamt$ was computed independently in~\cite{Chen:2023dsi}, and a perfect agreement was found regarding this piece. 

\end{acknowledgments}

\bibliographystyle{utphysMa}
\bibliography{TwbN3LO}

\providecommand{\href}[2]{#2}\begingroup\raggedright\begin{thebibliography}{10}

\bibitem{ILC:2013jhg}
{\bfseries ILC} , {\it {The International Linear Collider Technical Design
  Report - Volume 2: Physics}},
  [\href{http://arxiv.org/abs/1306.6352}{{\ttfamily arXiv:1306.6352}}]
  [\href{http://inspirehep.net/search?p=find+ILC:2013jhg}{{\ttfamily
  InSPIRE}}].

\bibitem{CLICdp:2018esa}
{\bfseries CLICdp} , H.~Abramowicz {\em et al.}, {\it {Top-Quark Physics at the
  CLIC Electron-Positron Linear Collider}},
  \href{http://dx.doi.org/10.1007/JHEP11(2019)003}{{\em JHEP} {\bfseries 11}
  (2019) 003} [\href{http://arxiv.org/abs/1807.02441}{{\ttfamily
  arXiv:1807.02441}}]
  [\href{http://inspirehep.net/search?p=find+CLICdp:2018esa}{{\ttfamily
  InSPIRE}}].

\bibitem{FCC:2018evy}
{\bfseries FCC} , A.~Abada {\em et al.}, {\it {FCC-ee: The Lepton Collider}:
  {Future Circular Collider Conceptual Design Report Volume 2}},
  \href{http://dx.doi.org/10.1140/epjst/e2019-900045-4}{{\em Eur. Phys. J. ST}
  {\bfseries 228} (2019) 261--623}
  [\href{http://inspirehep.net/search?p=find+FCC:2018evy}{{\ttfamily
  InSPIRE}}].

\bibitem{CEPCStudyGroup:2018ghi}
{\bfseries CEPC Study Group} , M.~Dong {\em et al.}, {\it {CEPC Conceptual
  Design Report: Volume 2 - Physics \& Detector}},
  [\href{http://arxiv.org/abs/1811.10545}{{\ttfamily arXiv:1811.10545}}]
  [\href{http://inspirehep.net/search?p=find+CEPCStudyGroup:2018ghi}{{\ttfamily
  InSPIRE}}].

\bibitem{ParticleDataGroup:2022pth}
{\bfseries Particle Data Group} , R.~L. Workman {\em et al.}, {\it {Review of
  Particle Physics}},  \href{http://dx.doi.org/10.1093/ptep/ptac097}{{\em PTEP}
  {\bfseries 2022} (2022) 083C01}
  [\href{http://inspirehep.net/search?p=find+ParticleDataGroup:2022pth}{{\ttfamily
  InSPIRE}}].

\bibitem{CMS:2014mxl}
{\bfseries CMS} , V.~Khachatryan {\em et al.}, {\it {Measurement of the ratio
  $\mathcal B(t \to Wb)/\mathcal B(t \to Wq)$ in pp collisions at $\sqrt{s}$ =
  8 TeV}},  \href{http://dx.doi.org/10.1016/j.physletb.2014.06.076}{{\em Phys.
  Lett. B} {\bfseries 736} (2014) 33--57}
  [\href{http://arxiv.org/abs/1404.2292}{{\ttfamily arXiv:1404.2292}}]
  [\href{http://inspirehep.net/search?p=find+CMS:2014mxl}{{\ttfamily
  InSPIRE}}].

\bibitem{Horiguchi:2013wra}
T.~Horiguchi, A.~Ishikawa, T.~Suehara, K.~Fujii, Y.~Sumino, Y.~Kiyo, and
  H.~Yamamoto, {\it {Study of top quark pair production near threshold at the
  ILC}},  [\href{http://arxiv.org/abs/1310.0563}{{\ttfamily arXiv:1310.0563}}]
  [\href{http://inspirehep.net/search?p=find+Horiguchi:2013wra}{{\ttfamily
  InSPIRE}}].

\bibitem{Baskakov:2018huw}
A.~Baskakov, E.~Boos, and L.~Dudko, {\it {Model independent top quark width
  measurement using a combination of resonant and nonresonant cross sections}},
   \href{http://dx.doi.org/10.1103/PhysRevD.98.116011}{{\em Phys. Rev. D}
  {\bfseries 98} (2018) 116011}
  [\href{http://arxiv.org/abs/1807.11193}{{\ttfamily arXiv:1807.11193}}]
  [\href{http://inspirehep.net/search?p=find+Baskakov:2018huw}{{\ttfamily
  InSPIRE}}].

\bibitem{Li:2022iav}
Z.~Li, X.~Sun, Y.~Fang, G.~Li, S.~Xin, S.~Wang, Y.~Wang, Y.~Zhang, H.~Zhang,
  and Z.~Liang, {\it {Top quark mass measurements at the $t\bar{t}$ threshold
  with CEPC}},  \href{http://dx.doi.org/10.1140/epjc/s10052-023-11421-1}{{\em
  Eur. Phys. J. C} {\bfseries 83} (2023) 269}
  [\href{http://arxiv.org/abs/2207.12177}{{\ttfamily arXiv:2207.12177}}]
  [\href{http://inspirehep.net/search?p=find+Li:2022iav}{{\ttfamily InSPIRE}}].
  [Erratum: Eur.Phys.J.C 83, 501 (2023)].

\bibitem{Jezabek:1988iv}
M.~Jezabek and J.~H. Kuhn, {\it {QCD Corrections to Semileptonic Decays of
  Heavy Quarks}},  \href{http://dx.doi.org/10.1016/0550-3213(89)90108-9}{{\em
  Nucl. Phys. B} {\bfseries 314} (1989) 1--6}
  [\href{http://inspirehep.net/search?p=find+Jezabek:1988iv}{{\ttfamily
  InSPIRE}}].

\bibitem{Czarnecki:1990kv}
A.~Czarnecki, {\it {QCD corrections to the decay t ---\ensuremath{>} W b in
  dimensional regularization}},
  \href{http://dx.doi.org/10.1016/0370-2693(90)90571-M}{{\em Phys. Lett. B}
  {\bfseries 252} (1990) 467--470}
  [\href{http://inspirehep.net/search?p=find+Czarnecki:1990kv}{{\ttfamily
  InSPIRE}}].

\bibitem{Li:1990qf}
C.~S. Li, R.~J. Oakes, and T.~C. Yuan, {\it {QCD corrections to $t \to W^{+}
  b$}},  \href{http://dx.doi.org/10.1103/PhysRevD.43.3759}{{\em Phys. Rev. D}
  {\bfseries 43} (1991) 3759--3762}
  [\href{http://inspirehep.net/search?p=find+Li:1990qf}{{\ttfamily InSPIRE}}].

\bibitem{Czarnecki:1998qc}
A.~Czarnecki and K.~Melnikov, {\it {Two loop QCD corrections to top quark
  width}},  \href{http://dx.doi.org/10.1016/S0550-3213(98)00844-X}{{\em Nucl.
  Phys. B} {\bfseries 544} (1999) 520--531}
  [\href{http://arxiv.org/abs/hep-ph/9806244}{{\ttfamily hep-ph/9806244}}]
  [\href{http://inspirehep.net/search?p=find+Czarnecki:1998qc}{{\ttfamily
  InSPIRE}}].

\bibitem{Chetyrkin:1999ju}
K.~G. Chetyrkin, R.~Harlander, T.~Seidensticker, and M.~Steinhauser, {\it
  {Second order QCD corrections to Gamma(t ---\ensuremath{>} W b)}},
  \href{http://dx.doi.org/10.1103/PhysRevD.60.114015}{{\em Phys. Rev. D}
  {\bfseries 60} (1999) 114015}
  [\href{http://arxiv.org/abs/hep-ph/9906273}{{\ttfamily hep-ph/9906273}}]
  [\href{http://inspirehep.net/search?p=find+Chetyrkin:1999ju}{{\ttfamily
  InSPIRE}}].

\bibitem{Fischer:2001gp}
M.~Fischer, S.~Groote, J.~G. Korner, and M.~C. Mauser, {\it {Complete angular
  analysis of polarized top decay at O(alpha($s$) )}},
  \href{http://dx.doi.org/10.1103/PhysRevD.65.054036}{{\em Phys. Rev. D}
  {\bfseries 65} (2002) 054036}
  [\href{http://arxiv.org/abs/hep-ph/0101322}{{\ttfamily hep-ph/0101322}}]
  [\href{http://inspirehep.net/search?p=find+Fischer:2001gp}{{\ttfamily
  InSPIRE}}].

\bibitem{Blokland:2004ye}
I.~R. Blokland, A.~Czarnecki, M.~Slusarczyk, and F.~Tkachov, {\it {Heavy to
  light decays with a two loop accuracy}},
  \href{http://dx.doi.org/10.1103/PhysRevLett.93.062001}{{\em Phys. Rev. Lett.}
  {\bfseries 93} (2004) 062001}
  [\href{http://arxiv.org/abs/hep-ph/0403221}{{\ttfamily hep-ph/0403221}}]
  [\href{http://inspirehep.net/search?p=find+Blokland:2004ye}{{\ttfamily
  InSPIRE}}].

\bibitem{Blokland:2005vq}
I.~R. Blokland, A.~Czarnecki, M.~Slusarczyk, and F.~Tkachov, {\it
  {Next-to-next-to-leading order calculations for heavy-to-light decays}},
  \href{http://dx.doi.org/10.1103/PhysRevD.71.054004}{{\em Phys. Rev. D}
  {\bfseries 71} (2005) 054004}
  [\href{http://arxiv.org/abs/hep-ph/0503039}{{\ttfamily hep-ph/0503039}}]
  [\href{http://inspirehep.net/search?p=find+Blokland:2005vq}{{\ttfamily
  InSPIRE}}]. [Erratum: Phys.Rev.D 79, 019901 (2009)].

\bibitem{Czarnecki:2010gb}
A.~Czarnecki, J.~G. Korner, and J.~H. Piclum, {\it {Helicity fractions of W
  bosons from top quark decays at NNLO in QCD}},
  \href{http://dx.doi.org/10.1103/PhysRevD.81.111503}{{\em Phys. Rev. D}
  {\bfseries 81} (2010) 111503}
  [\href{http://arxiv.org/abs/1005.2625}{{\ttfamily arXiv:1005.2625}}]
  [\href{http://inspirehep.net/search?p=find+Czarnecki:2010gb}{{\ttfamily
  InSPIRE}}].

\bibitem{Gao:2012ja}
J.~Gao, C.~S. Li, and H.~X. Zhu, {\it {Top Quark Decay at Next-to-Next-to
  Leading Order in QCD}},
  \href{http://dx.doi.org/10.1103/PhysRevLett.110.042001}{{\em Phys. Rev.
  Lett.} {\bfseries 110} (2013) 042001}
  [\href{http://arxiv.org/abs/1210.2808}{{\ttfamily arXiv:1210.2808}}]
  [\href{http://inspirehep.net/search?p=find+Gao:2012ja}{{\ttfamily InSPIRE}}].

\bibitem{Brucherseifer:2013iv}
M.~Brucherseifer, F.~Caola, and K.~Melnikov, {\it {$\mathcal O(\alpha_s^2)$
  corrections to fully-differential top quark decays}},
  \href{http://dx.doi.org/10.1007/JHEP04(2013)059}{{\em JHEP} {\bfseries 04}
  (2013) 059} [\href{http://arxiv.org/abs/1301.7133}{{\ttfamily
  arXiv:1301.7133}}]
  [\href{http://inspirehep.net/search?p=find+Brucherseifer:2013iv}{{\ttfamily
  InSPIRE}}].

\bibitem{Campbell:2020fhf}
J.~Campbell, T.~Neumann, and Z.~Sullivan, {\it {Single-top-quark production in
  the $t$-channel at NNLO}},
  \href{http://dx.doi.org/10.1007/JHEP02(2021)040}{{\em JHEP} {\bfseries 02}
  (2021) 040} [\href{http://arxiv.org/abs/2012.01574}{{\ttfamily
  arXiv:2012.01574}}]
  [\href{http://inspirehep.net/search?p=find+Campbell:2020fhf}{{\ttfamily
  InSPIRE}}].

\bibitem{Meng:2022htg}
R.-Q. Meng, S.-Q. Wang, T.~Sun, C.-Q. Luo, J.-M. Shen, and X.-G. Wu, {\it {QCD
  improved top-quark decay at next-to-next-to-leading order}},
  \href{http://dx.doi.org/10.1140/epjc/s10052-023-11224-4}{{\em Eur. Phys. J.
  C} {\bfseries 83} (2023) 59}
  [\href{http://arxiv.org/abs/2202.09978}{{\ttfamily arXiv:2202.09978}}]
  [\href{http://inspirehep.net/search?p=find+Meng:2022htg}{{\ttfamily
  InSPIRE}}].

\bibitem{Chen:2022wit}
L.-B. Chen, H.~T. Li, J.~Wang, and Y.~Wang, {\it {Analytic result for the
  top-quark width at next-to-next-to-leading order in QCD}},
  [\href{http://arxiv.org/abs/2212.06341}{{\ttfamily arXiv:2212.06341}}]
  [\href{http://inspirehep.net/search?p=find+Chen:2022wit}{{\ttfamily
  InSPIRE}}].

\bibitem{Beneke:1994qe}
M.~Beneke and V.~M. Braun, {\it {Naive nonAbelianization and resummation of
  fermion bubble chains}},
  \href{http://dx.doi.org/10.1016/0370-2693(95)00184-M}{{\em Phys. Lett. B}
  {\bfseries 348} (1995) 513--520}
  [\href{http://arxiv.org/abs/hep-ph/9411229}{{\ttfamily hep-ph/9411229}}]
  [\href{http://inspirehep.net/search?p=find+Beneke:1994qe}{{\ttfamily
  InSPIRE}}].

\bibitem{Smith:1996xz}
M.~C. Smith and S.~S. Willenbrock, {\it {Top quark pole mass}},
  \href{http://dx.doi.org/10.1103/PhysRevLett.79.3825}{{\em Phys. Rev. Lett.}
  {\bfseries 79} (1997) 3825--3828}
  [\href{http://arxiv.org/abs/hep-ph/9612329}{{\ttfamily hep-ph/9612329}}]
  [\href{http://inspirehep.net/search?p=find+Smith:1996xz}{{\ttfamily
  InSPIRE}}].

\bibitem{Beneke:1998ui}
M.~Beneke, {\it {Renormalons}},
  \href{http://dx.doi.org/10.1016/S0370-1573(98)00130-6}{{\em Phys. Rept.}
  {\bfseries 317} (1999) 1--142}
  [\href{http://arxiv.org/abs/hep-ph/9807443}{{\ttfamily hep-ph/9807443}}]
  [\href{http://inspirehep.net/search?p=find+Beneke:1998ui}{{\ttfamily
  InSPIRE}}].

\bibitem{Beneke:2016cbu}
M.~Beneke, P.~Marquard, P.~Nason, and M.~Steinhauser, {\it {On the ultimate
  uncertainty of the top quark pole mass}},
  \href{http://dx.doi.org/10.1016/j.physletb.2017.10.054}{{\em Phys. Lett. B}
  {\bfseries 775} (2017) 63--70}
  [\href{http://arxiv.org/abs/1605.03609}{{\ttfamily arXiv:1605.03609}}]
  [\href{http://inspirehep.net/search?p=find+Beneke:2016cbu}{{\ttfamily
  InSPIRE}}].

\bibitem{Hoang:2017btd}
A.~H. Hoang, C.~Lepenik, and M.~Preisser, {\it {On the Light Massive Flavor
  Dependence of the Large Order Asymptotic Behavior and the Ambiguity of the
  Pole Mass}},  \href{http://dx.doi.org/10.1007/JHEP09(2017)099}{{\em JHEP}
  {\bfseries 09} (2017) 099} [\href{http://arxiv.org/abs/1706.08526}{{\ttfamily
  arXiv:1706.08526}}]
  [\href{http://inspirehep.net/search?p=find+Hoang:2017btd}{{\ttfamily
  InSPIRE}}].

\bibitem{FerrarioRavasio:2018ubr}
S.~Ferrario~Ravasio, P.~Nason, and C.~Oleari, {\it {All-orders behaviour and
  renormalons in top-mass observables}},
  \href{http://dx.doi.org/10.1007/JHEP01(2019)203}{{\em JHEP} {\bfseries 01}
  (2019) 203} [\href{http://arxiv.org/abs/1810.10931}{{\ttfamily
  arXiv:1810.10931}}]
  [\href{http://inspirehep.net/search?p=find+FerrarioRavasio:2018ubr}{{\ttfamily
  InSPIRE}}].

\bibitem{ATLAS:2022rms}
{\bfseries ATLAS} , G.~Aad {\em et al.}, {\it {Measurement of the polarisation
  of W bosons produced in top-quark decays using dilepton events at s=13 TeV
  with the ATLAS experiment}},
  \href{http://dx.doi.org/10.1016/j.physletb.2023.137829}{{\em Phys. Lett. B}
  {\bfseries 843} (2023) 137829}
  [\href{http://arxiv.org/abs/2209.14903}{{\ttfamily arXiv:2209.14903}}]
  [\href{http://inspirehep.net/search?p=find+ATLAS:2022rms}{{\ttfamily
  InSPIRE}}].

\bibitem{Vos:2016til}
M.~Vos {\em et al.}, {\it {Top physics at high-energy lepton colliders}},
  [\href{http://arxiv.org/abs/1604.08122}{{\ttfamily arXiv:1604.08122}}]
  [\href{http://inspirehep.net/search?p=find+Vos:2016til}{{\ttfamily
  InSPIRE}}].

\bibitem{Schwienhorst:2022yqu}
K.~Agashe {\em et al.}, {\it {Report of the Topical Group on Top quark physics
  and heavy flavor production for Snowmass 2021}},
  [\href{http://arxiv.org/abs/2209.11267}{{\ttfamily arXiv:2209.11267}}]
  [\href{http://inspirehep.net/search?p=find+Schwienhorst:2022yqu}{{\ttfamily
  InSPIRE}}].

\bibitem{Fischer:2000kx}
M.~Fischer, S.~Groote, J.~G. Korner, and M.~C. Mauser, {\it {Longitudinal,
  transverse plus and transverse minus $W$ bosons in unpolarized top quark
  decays at O(alpha($s$) )}},
  \href{http://dx.doi.org/10.1103/PhysRevD.63.031501}{{\em Phys. Rev. D}
  {\bfseries 63} (2001) 031501}
  [\href{http://arxiv.org/abs/hep-ph/0011075}{{\ttfamily hep-ph/0011075}}]
  [\href{http://inspirehep.net/search?p=find+Fischer:2000kx}{{\ttfamily
  InSPIRE}}].

\bibitem{Chen:2019wyb}
L.~Chen, {\it {A prescription for projectors to compute helicity amplitudes in
  D dimensions}},
  \href{http://dx.doi.org/10.1140/epjc/s10052-021-09210-9}{{\em Eur. Phys. J.
  C} {\bfseries 81} (2021) 417}
  [\href{http://arxiv.org/abs/1904.00705}{{\ttfamily arXiv:1904.00705}}]
  [\href{http://inspirehep.net/search?p=find+Chen:2019wyb}{{\ttfamily
  InSPIRE}}].

\bibitem{Bardeen:1972vi}
W.~A. Bardeen, R.~Gastmans, and B.~E. Lautrup, {\it {Static quantities in
  Weinberg's model of weak and electromagnetic interactions}},
\href{http://dx.doi.org/10.1016/0550-3213(72)90218-0}{{\em Nucl. Phys.}
  {\bfseries B46} (1972) 319--331}
  [\href{http://inspirehep.net/search?p=find+Bardeen:1972vi}{{\ttfamily
  InSPIRE}}].

\bibitem{Chanowitz:1979zu}
M.~S. Chanowitz, M.~Furman, and I.~Hinchliffe, {\it {The Axial Current in
  Dimensional Regularization}},
\href{http://dx.doi.org/10.1016/0550-3213(79)90333-X}{{\em Nucl. Phys.}
  {\bfseries B159} (1979) 225--243}
  [\href{http://inspirehep.net/search?p=find+Chanowitz:1979zu}{{\ttfamily
  InSPIRE}}].

\bibitem{Gottlieb:1979ix}
S.~A. Gottlieb and J.~T. Donohue, {\it {The Axial Vector Current and
  Dimensional Regularization}},
\href{http://dx.doi.org/10.1103/PhysRevD.20.3378}{{\em Phys. Rev.} {\bfseries
  D20} (1979) 3378}
  [\href{http://inspirehep.net/search?p=find+Gottlieb:1979ix}{{\ttfamily
  InSPIRE}}].

\bibitem{Korner:1991sx}
J.~G. K{\"o}rner, D.~Kreimer, and K.~Schilcher, {\it {A Practicable $\gamma_5$
  scheme in dimensional regularization}},
\href{http://dx.doi.org/10.1007/BF01559471}{{\em Z. Phys.} {\bfseries C54}
  (1992) 503--512}
  [\href{http://inspirehep.net/search?p=find+Korner:1991sx}{{\ttfamily
  InSPIRE}}].

\bibitem{Chetyrkin:1981qh}
K.~G. Chetyrkin and F.~V. Tkachov, {\it {Integration by Parts: The Algorithm to
  Calculate beta Functions in 4 Loops}},
\href{http://dx.doi.org/10.1016/0550-3213(81)90199-1}{{\em Nucl. Phys.}
  {\bfseries B192} (1981) 159--204}
  [\href{http://inspirehep.net/search?p=find+Chetyrkin:1981qh}{{\ttfamily
  InSPIRE}}].

\bibitem{Guan:2024byi}
X.~Guan, X.~Liu, Y.-Q. Ma, and W.-H. Wu, {\it {Blade: A package for
  block-triangular form improved Feynman integrals decomposition}},
  [\href{http://arxiv.org/abs/2405.14621}{{\ttfamily arXiv:2405.14621}}]
  [\href{http://inspirehep.net/search?p=find+Guan:2024byi}{{\ttfamily
  InSPIRE}}].

\bibitem{Liu:2018dmc}
X.~Liu and Y.-Q. Ma, {\it {Determining arbitrary Feynman integrals by vacuum
  integrals}},  \href{http://dx.doi.org/10.1103/PhysRevD.99.071501}{{\em Phys.
  Rev. D} {\bfseries 99} (2019) 071501}
  [\href{http://arxiv.org/abs/1801.10523}{{\ttfamily arXiv:1801.10523}}]
  [\href{http://inspirehep.net/search?p=find+Liu:2018dmc}{{\ttfamily
  InSPIRE}}].

\bibitem{Guan:2019bcx}
X.~Guan, X.~Liu, and Y.-Q. Ma, {\it {Complete reduction of integrals in
  two-loop five-light-parton scattering amplitudes}},
  \href{http://dx.doi.org/10.1088/1674-1137/44/9/093106}{{\em Chin. Phys. C}
  {\bfseries 44} (2020) 093106}
  [\href{http://arxiv.org/abs/1912.09294}{{\ttfamily arXiv:1912.09294}}]
  [\href{http://inspirehep.net/search?p=find+Guan:2019bcx}{{\ttfamily
  InSPIRE}}].

\bibitem{Peraro:2019cjj}
T.~Peraro and L.~Tancredi, {\it {Physical projectors for multi-leg helicity
  amplitudes}},
\href{http://dx.doi.org/10.1007/JHEP07(2019)114}{{\em JHEP} {\bfseries 07}
  (2019) 114} [\href{http://arxiv.org/abs/1906.03298}{{\ttfamily
  arXiv:1906.03298}}]
  [\href{http://inspirehep.net/search?p=find+Peraro:2019cjj}{{\ttfamily
  InSPIRE}}].

\bibitem{Anastasiou:2004vj}
C.~Anastasiou and A.~Lazopoulos, {\it {Automatic integral reduction for higher
  order perturbative calculations}},
\href{http://dx.doi.org/10.1088/1126-6708/2004/07/046}{{\em JHEP} {\bfseries
  07} (2004) 046} [\href{http://arxiv.org/abs/hep-ph/0404258}{{\ttfamily
  hep-ph/0404258}}]
  [\href{http://inspirehep.net/search?p=find+Anastasiou:2004vj}{{\ttfamily
  InSPIRE}}].

\bibitem{Smirnov:2008iw}
A.~V. Smirnov, {\it {Algorithm FIRE -- Feynman Integral REduction}},
  \href{http://dx.doi.org/10.1088/1126-6708/2008/10/107}{{\em JHEP} {\bfseries
  10} (2008) 107} [\href{http://arxiv.org/abs/0807.3243}{{\ttfamily
  arXiv:0807.3243}}]
  [\href{http://inspirehep.net/search?p=find+Smirnov:2008iw}{{\ttfamily
  InSPIRE}}].

\bibitem{Smirnov:2013dia}
A.~V. Smirnov and V.~A. Smirnov, {\it {FIRE4, LiteRed and accompanying tools to
  solve integration by parts relations}},
  \href{http://dx.doi.org/10.1016/j.cpc.2013.06.016}{{\em Comput. Phys.
  Commun.} {\bfseries 184} (2013) 2820--2827}
  [\href{http://arxiv.org/abs/1302.5885}{{\ttfamily arXiv:1302.5885}}]
  [\href{http://inspirehep.net/search?p=find+Smirnov:2013dia}{{\ttfamily
  InSPIRE}}].

\bibitem{Smirnov:2014hma}
A.~V. Smirnov, {\it {FIRE5: a C++ implementation of Feynman Integral
  REduction}},
\href{http://dx.doi.org/10.1016/j.cpc.2014.11.024}{{\em Comput. Phys. Commun.}
  {\bfseries 189} (2015) 182--191}
  [\href{http://arxiv.org/abs/1408.2372}{{\ttfamily arXiv:1408.2372}}]
  [\href{http://inspirehep.net/search?p=find+Smirnov:2014hma}{{\ttfamily
  InSPIRE}}].

\bibitem{Smirnov:2019qkx}
A.~Smirnov and F.~Chuharev, {\it {FIRE6: Feynman Integral REduction with
  Modular Arithmetic}},  [\href{http://arxiv.org/abs/1901.07808}{{\ttfamily
  arXiv:1901.07808}}]
  [\href{http://inspirehep.net/search?p=find+Smirnov:2019qkx}{{\ttfamily
  InSPIRE}}].

\bibitem{Maierhofer:2017gsa}
P.~Maierh\"ofer, J.~Usovitsch, and P.~Uwer, {\it {Kira\textemdash{}A Feynman
  integral reduction program}},
  \href{http://dx.doi.org/10.1016/j.cpc.2018.04.012}{{\em Comput. Phys.
  Commun.} {\bfseries 230} (2018) 99--112}
  [\href{http://arxiv.org/abs/1705.05610}{{\ttfamily arXiv:1705.05610}}]
  [\href{http://inspirehep.net/search?p=find+Maierhofer:2017gsa}{{\ttfamily
  InSPIRE}}].

\bibitem{Maierhofer:2018gpa}
P.~Maierhöfer and J.~Usovitsch,
{\it {Kira 1.2 Release Notes}},
  [\href{http://arxiv.org/abs/1812.01491}{{\ttfamily arXiv:1812.01491}}]
  [\href{http://inspirehep.net/search?p=find+Maierhofer:2018gpa}{{\ttfamily
  InSPIRE}}].

\bibitem{Klappert:2020nbg}
J.~Klappert, F.~Lange, P.~Maierh\"ofer, and J.~Usovitsch, {\it {Integral
  reduction with Kira 2.0 and finite field methods}},
  \href{http://dx.doi.org/10.1016/j.cpc.2021.108024}{{\em Comput. Phys.
  Commun.} {\bfseries 266} (2021) 108024}
  [\href{http://arxiv.org/abs/2008.06494}{{\ttfamily arXiv:2008.06494}}]
  [\href{http://inspirehep.net/search?p=find+Klappert:2020nbg}{{\ttfamily
  InSPIRE}}].

\bibitem{Lee:2012cn}
R.~N. Lee,
{\it {Presenting LiteRed: a tool for the Loop InTEgrals REDuction}},
  [\href{http://arxiv.org/abs/1212.2685}{{\ttfamily arXiv:1212.2685}}]
  [\href{http://inspirehep.net/search?p=find+Lee:2012cn}{{\ttfamily InSPIRE}}].

\bibitem{Lee:2013mka}
R.~N. Lee, {\it {LiteRed 1.4: a powerful tool for reduction of multiloop
  integrals}},  \href{http://dx.doi.org/10.1088/1742-6596/523/1/012059}{{\em J.
  Phys. Conf. Ser.} {\bfseries 523} (2014) 012059}
  [\href{http://arxiv.org/abs/1310.1145}{{\ttfamily arXiv:1310.1145}}]
  [\href{http://inspirehep.net/search?p=find+Lee:2013mka}{{\ttfamily
  InSPIRE}}].

\bibitem{Studerus:2009ye}
C.~Studerus, {\it {Reduze-Feynman Integral Reduction in C++}},
\href{http://dx.doi.org/10.1016/j.cpc.2010.03.012}{{\em Comput. Phys. Commun.}
  {\bfseries 181} (2010) 1293--1300}
  [\href{http://arxiv.org/abs/0912.2546}{{\ttfamily arXiv:0912.2546}}]
  [\href{http://inspirehep.net/search?p=find+Studerus:2009ye}{{\ttfamily
  InSPIRE}}].

\bibitem{vonManteuffel:2012np}
A.~von Manteuffel and C.~Studerus,
{\it {Reduze 2 - Distributed Feynman Integral Reduction}},
  [\href{http://arxiv.org/abs/1201.4330}{{\ttfamily arXiv:1201.4330}}]
  [\href{http://inspirehep.net/search?p=find+vonManteuffel:2012np}{{\ttfamily
  InSPIRE}}].

\bibitem{Wu:2023upw}
Z.~Wu, J.~Boehm, R.~Ma, H.~Xu, and Y.~Zhang, {\it {NeatIBP 1.0, A package
  generating small-size integration-by-parts relations for Feynman integrals}},
   [\href{http://arxiv.org/abs/2305.08783}{{\ttfamily arXiv:2305.08783}}]
  [\href{http://inspirehep.net/search?p=find+Wu:2023upw}{{\ttfamily InSPIRE}}].

\bibitem{Kotikov:1990kg}
A.~V. Kotikov, {\it {Differential equations method: New technique for massive
  Feynman diagrams calculation}},
  \href{http://dx.doi.org/10.1016/0370-2693(91)90413-K}{{\em Phys. Lett. B}
  {\bfseries 254} (1991) 158--164}
  [\href{http://inspirehep.net/search?p=find+Kotikov:1990kg}{{\ttfamily
  InSPIRE}}].

\bibitem{Remiddi:1997ny}
E.~Remiddi, {\it {Differential equations for Feynman graph amplitudes}},  {\em
  Nuovo Cim. A} {\bfseries 110} (1997) 1435--1452
  [\href{http://arxiv.org/abs/hep-th/9711188}{{\ttfamily hep-th/9711188}}]
  [\href{http://inspirehep.net/search?p=find+Remiddi:1997ny}{{\ttfamily
  InSPIRE}}].

\bibitem{Caffo:2008aw}
M.~Caffo, H.~Czyz, M.~Gunia, and E.~Remiddi, {\it {BOKASUN: A Fast and precise
  numerical program to calculate the Master Integrals of the two-loop sunrise
  diagrams}},  \href{http://dx.doi.org/10.1016/j.cpc.2008.10.011}{{\em Comput.
  Phys. Commun.} {\bfseries 180} (2009) 427--430}
  [\href{http://arxiv.org/abs/0807.1959}{{\ttfamily arXiv:0807.1959}}]
  [\href{http://inspirehep.net/search?p=find+Caffo:2008aw}{{\ttfamily
  InSPIRE}}].

\bibitem{Czakon:2008zk}
M.~Czakon, {\it {Tops from Light Quarks: Full Mass Dependence at Two-Loops in
  QCD}},  \href{http://dx.doi.org/10.1016/j.physletb.2008.05.028}{{\em Phys.
  Lett. B} {\bfseries 664} (2008) 307--314}
  [\href{http://arxiv.org/abs/0803.1400}{{\ttfamily arXiv:0803.1400}}]
  [\href{http://inspirehep.net/search?p=find+Czakon:2008zk}{{\ttfamily
  InSPIRE}}].

\bibitem{Lee:2014ioa}
R.~N. Lee, {\it {Reducing differential equations for multiloop master
  integrals}},  \href{http://dx.doi.org/10.1007/JHEP04(2015)108}{{\em JHEP}
  {\bfseries 04} (2015) 108} [\href{http://arxiv.org/abs/1411.0911}{{\ttfamily
  arXiv:1411.0911}}]
  [\href{http://inspirehep.net/search?p=find+Lee:2014ioa}{{\ttfamily
  InSPIRE}}].

\bibitem{Moriello:2019yhu}
F.~Moriello, {\it {Generalised power series expansions for the elliptic planar
  families of Higgs + jet production at two loops}},
  \href{http://dx.doi.org/10.1007/JHEP01(2020)150}{{\em JHEP} {\bfseries 01}
  (2020) 150} [\href{http://arxiv.org/abs/1907.13234}{{\ttfamily
  arXiv:1907.13234}}]
  [\href{http://inspirehep.net/search?p=find+Moriello:2019yhu}{{\ttfamily
  InSPIRE}}].

\bibitem{Hidding:2020ytt}
M.~Hidding, {\it {DiffExp, a Mathematica package for computing Feynman
  integrals in terms of one-dimensional series expansions}},
  \href{http://dx.doi.org/10.1016/j.cpc.2021.108125}{{\em Comput. Phys.
  Commun.} {\bfseries 269} (2021) 108125}
  [\href{http://arxiv.org/abs/2006.05510}{{\ttfamily arXiv:2006.05510}}]
  [\href{http://inspirehep.net/search?p=find+Hidding:2020ytt}{{\ttfamily
  InSPIRE}}].

\bibitem{Armadillo:2022ugh}
T.~Armadillo, R.~Bonciani, S.~Devoto, N.~Rana, and A.~Vicini, {\it {Evaluation
  of Feynman integrals with arbitrary complex masses via series expansions}},
  \href{http://dx.doi.org/10.1016/j.cpc.2022.108545}{{\em Comput. Phys.
  Commun.} {\bfseries 282} (2023) 108545}
  [\href{http://arxiv.org/abs/2205.03345}{{\ttfamily arXiv:2205.03345}}]
  [\href{http://inspirehep.net/search?p=find+Armadillo:2022ugh}{{\ttfamily
  InSPIRE}}].

\bibitem{Liu:2017jxz}
X.~Liu, Y.-Q. Ma, and C.-Y. Wang, {\it {A Systematic and Efficient Method to
  Compute Multi-loop Master Integrals}},
  \href{http://dx.doi.org/10.1016/j.physletb.2018.02.026}{{\em Phys. Lett. B}
  {\bfseries 779} (2018) 353--357}
  [\href{http://arxiv.org/abs/1711.09572}{{\ttfamily arXiv:1711.09572}}]
  [\href{http://inspirehep.net/search?p=find+Liu:2017jxz}{{\ttfamily
  InSPIRE}}].

\bibitem{Liu:2020kpc}
X.~Liu, Y.-Q. Ma, W.~Tao, and P.~Zhang, {\it {Calculation of Feynman loop
  integration and phase-space integration via auxiliary mass flow}},
  \href{http://dx.doi.org/10.1088/1674-1137/abc538}{{\em Chin. Phys. C}
  {\bfseries 45} (2021) 013115}
  [\href{http://arxiv.org/abs/2009.07987}{{\ttfamily arXiv:2009.07987}}]
  [\href{http://inspirehep.net/search?p=find+Liu:2020kpc}{{\ttfamily
  InSPIRE}}].

\bibitem{Liu:2021wks}
X.~Liu and Y.-Q. Ma, {\it {Multiloop corrections for collider processes using
  auxiliary mass flow}},
  \href{http://dx.doi.org/10.1103/PhysRevD.105.L051503}{{\em Phys. Rev. D}
  {\bfseries 105} (2022) L051503}
  [\href{http://arxiv.org/abs/2107.01864}{{\ttfamily arXiv:2107.01864}}]
  [\href{http://inspirehep.net/search?p=find+Liu:2021wks}{{\ttfamily
  InSPIRE}}].

\bibitem{Liu:2022mfb}
Z.-F. Liu and Y.-Q. Ma, {\it {Determining Feynman Integrals with Only Input
  from Linear Algebra}},
  \href{http://dx.doi.org/10.1103/PhysRevLett.129.222001}{{\em Phys. Rev.
  Lett.} {\bfseries 129} (2022) 222001}
  [\href{http://arxiv.org/abs/2201.11637}{{\ttfamily arXiv:2201.11637}}]
  [\href{http://inspirehep.net/search?p=find+Liu:2022mfb}{{\ttfamily
  InSPIRE}}].

\bibitem{Liu:2022chg}
X.~Liu and Y.-Q. Ma, {\it {AMFlow: A Mathematica package for Feynman integrals
  computation via auxiliary mass flow}},
  \href{http://dx.doi.org/10.1016/j.cpc.2022.108565}{{\em Comput. Phys.
  Commun.} {\bfseries 283} (2023) 108565}
  [\href{http://arxiv.org/abs/2201.11669}{{\ttfamily arXiv:2201.11669}}]
  [\href{http://inspirehep.net/search?p=find+Liu:2022chg}{{\ttfamily
  InSPIRE}}].

\bibitem{Anastasiou:2002yz}
C.~Anastasiou and K.~Melnikov, {\it {Higgs boson production at hadron colliders
  in NNLO QCD}},  \href{http://dx.doi.org/10.1016/S0550-3213(02)00837-4}{{\em
  Nucl. Phys. B} {\bfseries 646} (2002) 220--256}
  [\href{http://arxiv.org/abs/hep-ph/0207004}{{\ttfamily hep-ph/0207004}}]
  [\href{http://inspirehep.net/search?p=find+Anastasiou:2002yz}{{\ttfamily
  InSPIRE}}].

\bibitem{Anastasiou:2002qz}
C.~Anastasiou, L.~J. Dixon, and K.~Melnikov, {\it {NLO Higgs boson rapidity
  distributions at hadron colliders}},
  \href{http://dx.doi.org/10.1016/S0920-5632(03)80168-8}{{\em Nucl. Phys. B
  Proc. Suppl.} {\bfseries 116} (2003) 193--197}
  [\href{http://arxiv.org/abs/hep-ph/0211141}{{\ttfamily hep-ph/0211141}}]
  [\href{http://inspirehep.net/search?p=find+Anastasiou:2002qz}{{\ttfamily
  InSPIRE}}].

\bibitem{Anastasiou:2003yy}
C.~Anastasiou, L.~J. Dixon, K.~Melnikov, and F.~Petriello, {\it {Dilepton
  rapidity distribution in the Drell-Yan process at NNLO in QCD}},
  \href{http://dx.doi.org/10.1103/PhysRevLett.91.182002}{{\em Phys. Rev. Lett.}
  {\bfseries 91} (2003) 182002}
  [\href{http://arxiv.org/abs/hep-ph/0306192}{{\ttfamily hep-ph/0306192}}]
  [\href{http://inspirehep.net/search?p=find+Anastasiou:2003yy}{{\ttfamily
  InSPIRE}}].

\bibitem{vanRitbergen:1997va}
T.~van Ritbergen, J.~Vermaseren, and S.~Larin, {\it {The Four loop beta
  function in quantum chromodynamics}},
  \href{http://dx.doi.org/10.1016/S0370-2693(97)00370-5}{{\em Phys. Lett. B}
  {\bfseries 400} (1997) 379--384}
  [\href{http://arxiv.org/abs/hep-ph/9701390}{{\ttfamily hep-ph/9701390}}]
  [\href{http://inspirehep.net/search?p=find+vanRitbergen:1997va}{{\ttfamily
  InSPIRE}}].

\bibitem{Chetyrkin:1999pq}
K.~Chetyrkin and A.~Retey, {\it {Renormalization and running of quark mass and
  field in the regularization invariant and MS-bar schemes at three loops and
  four loops}},  \href{http://dx.doi.org/10.1016/S0550-3213(00)00331-X}{{\em
  Nucl. Phys. B} {\bfseries 583} (2000) 3--34}
  [\href{http://arxiv.org/abs/hep-ph/9910332}{{\ttfamily hep-ph/9910332}}]
  [\href{http://inspirehep.net/search?p=find+Chetyrkin:1999pq}{{\ttfamily
  InSPIRE}}].

\bibitem{Czakon:2004bu}
M.~Czakon, {\it {The Four-loop QCD beta-function and anomalous dimensions}},
  \href{http://dx.doi.org/10.1016/j.nuclphysb.2005.01.012}{{\em Nucl. Phys. B}
  {\bfseries 710} (2005) 485--498}
  [\href{http://arxiv.org/abs/hep-ph/0411261}{{\ttfamily hep-ph/0411261}}]
  [\href{http://inspirehep.net/search?p=find+Czakon:2004bu}{{\ttfamily
  InSPIRE}}].

\bibitem{Chetyrkin:2004mf}
K.~Chetyrkin, {\it {Four-loop renormalization of QCD: Full set of
  renormalization constants and anomalous dimensions}},
  \href{http://dx.doi.org/10.1016/j.nuclphysb.2005.01.011}{{\em Nucl. Phys. B}
  {\bfseries 710} (2005) 499--510}
  [\href{http://arxiv.org/abs/hep-ph/0405193}{{\ttfamily hep-ph/0405193}}]
  [\href{http://inspirehep.net/search?p=find+Chetyrkin:2004mf}{{\ttfamily
  InSPIRE}}].

\bibitem{Melnikov:2000qh}
K.~Melnikov and T.~v. Ritbergen, {\it {The Three loop relation between the
  MS-bar and the pole quark masses}},
  \href{http://dx.doi.org/10.1016/S0370-2693(00)00507-4}{{\em Phys. Lett. B}
  {\bfseries 482} (2000) 99--108}
  [\href{http://arxiv.org/abs/hep-ph/9912391}{{\ttfamily hep-ph/9912391}}]
  [\href{http://inspirehep.net/search?p=find+Melnikov:2000qh}{{\ttfamily
  InSPIRE}}].

\bibitem{Aguilar-Saavedra:2015yza}
J.~A. Aguilar-Saavedra and J.~Bernabeu, {\it {Breaking down the entire W boson
  spin observables from its decay}},
  \href{http://dx.doi.org/10.1103/PhysRevD.93.011301}{{\em Phys. Rev. D}
  {\bfseries 93} (2016) 011301}
  [\href{http://arxiv.org/abs/1508.04592}{{\ttfamily arXiv:1508.04592}}]
  [\href{http://inspirehep.net/search?p=find+Aguilar-Saavedra:2015yza}{{\ttfamily
  InSPIRE}}].

\bibitem{Jezabek:1988ja}
M.~Jezabek and J.~H. Kuhn, {\it {Lepton Spectra from Heavy Quark Decay}},
  \href{http://dx.doi.org/10.1016/0550-3213(89)90209-5}{{\em Nucl. Phys. B}
  {\bfseries 320} (1989) 20--44}
  [\href{http://inspirehep.net/search?p=find+Jezabek:1988ja}{{\ttfamily
  InSPIRE}}].

\bibitem{Denner:1990cpz}
A.~Denner and T.~Sack, {\it {The W-boson width}},
  \href{http://dx.doi.org/10.1007/BF01560267}{{\em Z. Phys. C} {\bfseries 46}
  (1990) 653--663}
  [\href{http://inspirehep.net/search?p=find+Denner:1990cpz}{{\ttfamily
  InSPIRE}}].

\bibitem{Denner:1990ns}
A.~Denner and T.~Sack, {\it {The Top width}},
  \href{http://dx.doi.org/10.1016/0550-3213(91)90530-B}{{\em Nucl. Phys. B}
  {\bfseries 358} (1991) 46--58}
  [\href{http://inspirehep.net/search?p=find+Denner:1990ns}{{\ttfamily
  InSPIRE}}].

\bibitem{Eilam:1991iz}
G.~Eilam, R.~R. Mendel, R.~Migneron, and A.~Soni, {\it {Radiative corrections
  to top quark decay}},
  \href{http://dx.doi.org/10.1103/PhysRevLett.66.3105}{{\em Phys. Rev. Lett.}
  {\bfseries 66} (1991) 3105--3108}
  [\href{http://inspirehep.net/search?p=find+Eilam:1991iz}{{\ttfamily
  InSPIRE}}].

\bibitem{Do:2002ky}
H.~S. Do, S.~Groote, J.~G. Korner, and M.~C. Mauser, {\it {Electroweak and
  finite width corrections to top quark decays into transverse and longitudinal
  $W$ bosons}},  \href{http://dx.doi.org/10.1103/PhysRevD.67.091501}{{\em Phys.
  Rev. D} {\bfseries 67} (2003) 091501}
  [\href{http://arxiv.org/abs/hep-ph/0209185}{{\ttfamily hep-ph/0209185}}]
  [\href{http://inspirehep.net/search?p=find+Do:2002ky}{{\ttfamily InSPIRE}}].

\bibitem{Basso:2015gca}
L.~Basso, S.~Dittmaier, A.~Huss, and L.~Oggero, {\it {Techniques for the
  treatment of IR divergences in decay processes at NLO and application to the
  top-quark decay}},
  \href{http://dx.doi.org/10.1140/epjc/s10052-016-3878-2}{{\em Eur. Phys. J. C}
  {\bfseries 76} (2016) 56} [\href{http://arxiv.org/abs/1507.04676}{{\ttfamily
  arXiv:1507.04676}}]
  [\href{http://inspirehep.net/search?p=find+Basso:2015gca}{{\ttfamily
  InSPIRE}}].

\bibitem{Datta:2023otd}
S.~Datta, N.~Rana, V.~Ravindran, and R.~Sarkar, {\it {Three loop QCD
  corrections to the heavy-light form factors in the color-planar limit}},
  \href{http://dx.doi.org/10.1007/JHEP12(2023)001}{{\em JHEP} {\bfseries 12}
  (2023) 001} [\href{http://arxiv.org/abs/2308.12169}{{\ttfamily
  arXiv:2308.12169}}]
  [\href{http://inspirehep.net/search?p=find+Datta:2023otd}{{\ttfamily
  InSPIRE}}].

\bibitem{Chen:2023dsi}
L.-B. Chen, H.~T. Li, Z.~Li, J.~Wang, Y.~Wang, and Q.-f. Wu, {\it {Analytic
  third-order QCD corrections to top-quark and semileptonic b{\textrightarrow}u
  decays}},  \href{http://dx.doi.org/10.1103/PhysRevD.109.L071503}{{\em Phys.
  Rev. D} {\bfseries 109} (2024) L071503}
  [\href{http://arxiv.org/abs/2309.00762}{{\ttfamily arXiv:2309.00762}}]
  [\href{http://inspirehep.net/search?p=find+Chen:2023dsi}{{\ttfamily
  InSPIRE}}].

\end{thebibliography}\endgroup

\end{document}